# Statistical Methods for Determining Optimal Rifle Cartridge Dimensions

Steven Matthew Anderson[a], Shahar Boneh, Nels Grevstad
Department of Mathematics, Metropolitan State College of Denver, Denver, CO 80217-3362
[a]Also a statistician at Qwest Communication, Denver, CO, 80202

**Abstract**
We have designed and carried out a statistical study to determine the optimal cartridge dimensions for a Savage 10FLP law enforcement grade rifle. Optimal performance is defined as minimal group diameter. A full factorial block design with two main factors and one blocking factor was used. The two main factors were bullet seating depth and powder charge. The experimental units were individual shots taken from a bench-rest position and fired into separate targets. Additionally, thirteen covariates describing various cartridge dimensions were recorded. The data analysis includes ANOVA and ANCOVA. We will describe the experiment, the analysis, and some results.

## 1. Introduction

When a rifle cartridge is discharged, the bullet is accelerated down the bore of the barrel by pressure caused by the combustion of the explosive propellant contained in the cartridge. As the bullet is traveling down the barrel of a modern rifle it is spun by groves in the bore known as rifling. This acceleration and rotation of the mass causes a three dimensional vibration along the length of the barrel which reminds one of a corkscrew. Thus, depending on the last vibration node, the muzzle of the barrel travels in a cone, whose axis was the axis of the barrel before the ignition of the powder. This subsequent motion of the rifle barrel causes the initial direction of the shot to lie in a cone about the desired point of impact (i.e. the bulls-eye) (Mallock 1901).

The precision of a rifle is dependent on ones ability to tune this vibration so that deviation of the rifle muzzle is minimized from the line of sight for every shot. It is a characteristic that is unique to each individual firearm.

Both the cartridge dimensions and the physical characteristics of the rifle govern the barrels vibration pattern. Since a rifle's chamber dimensions, center of mass and the characteristics of the barrel/chamber coupling are difficult to adjust, it is easier to assume this as a fixed system and manipulate the dimensions of the cartridge. The rifle cartridge dimensions can be controlled by hand-loading the ammunition. Hand-loading is the act of taking the cartridge components and manufacturing the cartridges to the desired dimensions.

Surprisingly, a review of the literature found no systematic method for determining these optimal characteristics using design of experiment techniques. Competitive marksmen use their intuition, experience and trial and error to determine the optimal cartridge dimensions for their rifle. Design of experiment methodologies do not appear to be used. Therefore, we set out do design and conduct an experiment to determine such optimization.

## 2. Experimental Variables

### 2.1 The response variable
The measure of precision is a characteristic of a group of shots from cartridges with a particular response level. There are various methods to measure the precision, such as the measuring the extreme spread of the group (Cacoullos and Decicco, 1967), or measuring the group mean radius (see Wheeler 2007; Johnson 2001). Although either of these measures is a valid method for determining the precision, we chose to study the group mean radius for this experiment. The group mean radius (MR) is the average Euclidean distance from the shot $(x_i, y_i)$ to the group center $(\bar{x}, \bar{y})$. This is determined from the equation,

$$MR = \frac{\sum_{i=1}^{N}\sqrt{(x_i - \bar{x})^2 + (y_i - \bar{y})^2}}{N}$$

Both the extreme spread and the MR are easy to interpret. However the MR has the advantage of allowing one to treat each shot as an experimental unit as apposed to having one value per experimental level. Thus, allowing more degrees of freedom in the ANOVA or ANCOVA calculations. The following diagram illustrates the measuring of the mean radius of the group.

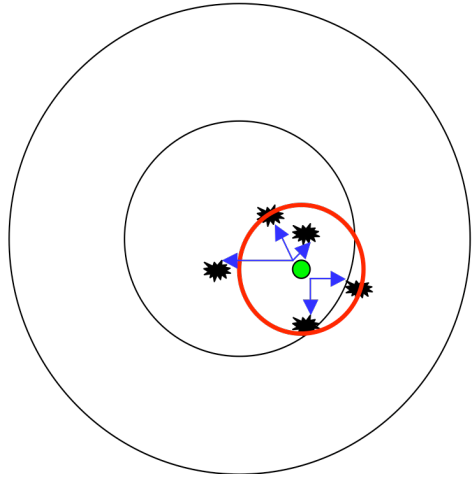

**Figure 1:** The group mean radius is the average of the individual Euclidean distances from the group center.

## 2.2 Main Factors
This was a full factorial design with two main factors. The main factors chosen for this experiment were seating depth and powder charge.

Seating depth is a measure of the distance from the bearing surface of the bullet to the start of the rifling in the barrel. The seating depths ranged from 0.005 inches to 0.030 inches, incremented by 0.005 inches, for a total of six factor levels.

The powder charge is the amount of propellant used in the cartridge, measured by its weight. The powder charges ranged from 25.3 grains to 26.2 grains, incremented by 0.1 grains, for a total of ten factor levels.

From a physical point of view, each experimental level determines an initial acceleration for the bullet passing through the barrel, when we assume that all other variables affecting the barrel vibration are kept constant.

## 2.3 Covariates
A 29-step case preparation process was performed. Some of the steps included cleaning, trimming to uniform length, machining the necks diameters and making the primer pockets and flash holes uniform. The Hornady 55 grain VMAX bullets were sorted by weight and length measured from base to bearing surface. The Winchester small rifle primers were all from the same manufacture lot to minimize variation. Even though extensive case preparation and sorting of components was used to minimize variation in these covariates, some differences from unit to unit were present.

Therefore, in addition to the two main factors we considered thirteen covariates that describe the specific dimensions of the cartridge and hopefully should account for some of the unit to unit variation.

All geometrical dimensions were measured by a digital calliper and recorded in thousands-of-an-inch. Weight measurements were performed with the use of a digital scale accurate to 0.1 grains. The covariates used in this study are listed in the Table 1 and shown in Figure 2 below. Table 2 summarizes the descriptive statistics of the measured covariate values.

| Table 1. Description of Covariates. | | |
|---|---|---|
| Covariate | Unit of Measure | Description |
| Case Length | Inches | Overall case length from head of case to mouth of case neck. Measured in inches |
| Neck Inner Diameter | Inches | Inner diameter of the case neck. |
| Neck Outer Diameter | Inches | Outer diameter of the case neck. |
| Neck Thickness | Inches | Thickness of the case neck as measured from the neck inner diameter to the neck outer diameter. |
| Head Space | Inches | Headspace is defined as the distance from the breech face of the bolt to the part of the chamber that stops the forward movement of the cartridge case. For a rimless bottlenecked cartridge, the headspace is the distance between breech face and the datum point on the shoulder defined by SAAMI. The measurement is the deviation from the SAAMI standard in thousands of an inch. |
| Primer Pocket Depth | Inches | Distance from the case head to the face of the primer pocket. |
| Primer Pocket Diameter | Inches | Diameter of the primer pocket. |
| Case Weight | Grains | Weight in grains of the unloaded case. |
| Case Volume | Grains | Measurement of the inner volume of the unloaded case. This was done by filling the unloaded case with 0.1 mm glass beads and weighing the media to obtain a volume by weight measurement. |
| Primer Weight | Grains | Weight of the primer before being inserted into the primer pocket. |
| Bullet Overall Length | Inches | Length of bullet as measured from tip to base. |
| Bullet Weight | Grains | Weight of the bullet before being inserted into the case neck. |
| Case Mouth Square | | This is a dichotomous variable that defines if the case neck was cut square (1="yes", 0="no"). Some of the case necks were cut unevenly due to a warped case neck die. |

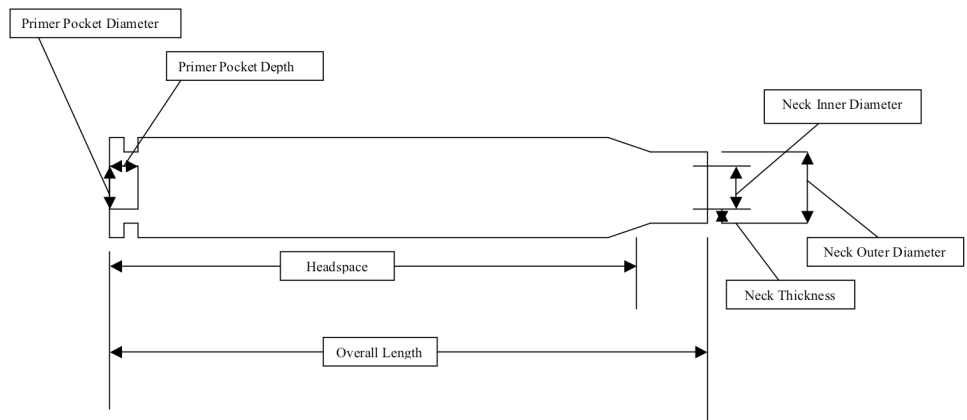

**Figure 2:** Covariates Used to Define the Cartridge Case Dimensions.

| Table 2. Descriptive Statistics of Covariates (N=380). | | | | | |
|---|---|---|---|---|---|
| Covariate | Mean | Std Dev | Minimum | Maximum | Median |
| Case Length | 1.7526 | 0.0044 | 1.7420 | 1.7630 | 1.7540 |
| Neck Inner Diameter | 0.2200 | 0.0005 | 0.2190 | 0.2210 | 0.2200 |
| Neck Outer Diameter | 0.2414 | 0.0006 | 0.2380 | 0.2430 | 0.2415 |
| Neck Thickness | 0.0114 | 0.0005 | 0.0090 | 0.0125 | 0.0115 |
| Head Space | -8.4303 | 1.8566 | -12.5000 | -2.0000 | -8.0000 |
| Primer Pocket Depth | 0.1190 | 0.0012 | 0.1120 | 0.1225 | 0.1190 |
| Primer Pocket Diameter | 0.1710 | 0.0005 | 0.1695 | 0.1720 | 0.1710 |
| Case Weight | 92.9526 | 0.7691 | 90.6000 | 95.3000 | 92.9000 |
| Case Mouth Square | 0.7316 | 0.4437 | 0.0000 | 1.0000 | 1.0000 |
| Case Volume | 45.8341 | 0.5118 | 44.3309 | 47.3622 | 45.7386 |
| Primer Weight | 3.2650 | 0.0478 | 3.2000 | 3.3000 | 3.3000 |
| Bullet Overall Length | 0.8107 | 0.0010 | 0.8080 | 0.8130 | 0.8110 |
| Bullet Weight | 55.0195 | 0.0711 | 54.8000 | 55.2000 | 55.0000 |

# 3. Experimental Details

## 3.1 Experimental Design
In this experiment we used a full factorial design with two main factors, 13 covariates and one blocking factor. Since there are 6 levels for seating depth and 10 levels for powder charge we had a total of 60 experimental levels. The cases were prepared in four lots of 100 cases each for a total of 400 experimental units. As we prepared the cases we found statistically significant differences in the covariates between lots. For this reason, we chose to use the lot as a blocking factor the model.

We randomized across the main factors within each lot in the following manner. We first randomly assigned a case from each lot to each experimental level for a total of 60 cases per lot. We then took the remaining 40 cases per lot and randomly assigned them an experimental level. Each level ended up having between five and eight cases. Each case was numbered with a permanent marker to identify it.

## 3.2 Equipment and Experimental Set-up
The rifle used was a Savage 10FLP chambered in .223 Remington. The distance from the rifle to the target was fixed at 100 yards. Each shot was fired into a separate target so that we could match each shot with a specific result. The barrel was cleaned after every 20 shots, except for the last 40 due to rain.

The shots were fired from a sitting bench rest position. The rifle was supported on a bench platform with the shooter sitting beside it. A tripod with a shot bag supported the forearm of the rifle and a contoured shot bag supported the rear of the rifle stock. We set up a chronograph 10 feet from the muzzle of the rifle to measure the velocity of each bullet. However, in order for the chronograph to function properly, the bullet must pass directly lengthwise through the center of the apparatus. Since the targets were positioned over a wide horizontal range on the backstop, the chronograph failed to record the shots fired at the extreme targets. As a result, only 50% of the velocities were actually recorded. The shooting was performed over a two-day period. The temperature during the first day ranged from 60-74 degrees F. The temperature on the second day ranged from 50-55 degrees F.

Figure 3 below describes the experimental set up and shows the shooting position.

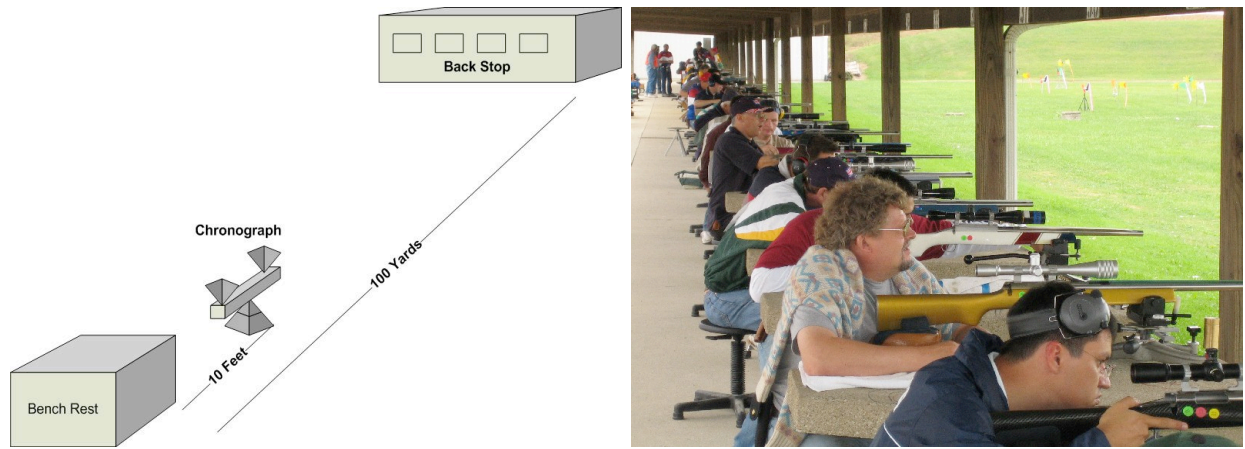

**Figure 3:** Experimental Set-Up and Shooting Position.

## 3.3 Nuisance & Bad Data Points

A 10-20 mph crosswind was present on both days with intermittent rain. If the wind was constant, the deviations from the group means would not be affected. However, it was not the case. An attempt was made to mark shots that might have been affected by wind gusts, but this was abandoned after several hours of shooting due to fatigue and time constraints. Ten shots were invalidated due to human error. Specifically, the shooter could feel the rifle pulling either to the right or left when he pulled the trigger. These 10 shots were later confirmed as being "bad" based on their large deviations from their respective group means.

After the shooting was completed, an examination of the rifle revealed that the bore had a large amount of copper fouling. This is caused by copper from the bullet jacket rubbing off as the bullet travels down the barrel. Normal cleaning procedure should have removed this copper from the barrel but apparently it did not. Copper fouling might cause a bullet to deform as it travels down the barrel and become aerodynamically unstable as it travels though the air. Indeed, after the shooting was finished, it was determined that four shots tumbled (i.e. hit the target sideways) which we attribute to this phenomenon.

The distance from the bulls-eye was measured over time and the variation was determined to be constant over the two-day shooting period. Figure 4 shows a time plot of the distances from the bulls-eye in the time order in which the shots were fired. Therefore, one can assume the copper fouling problem existed before the experiment started and deviations induced by this problem remained constant. The four tumbled bullets were invalidated from the results.

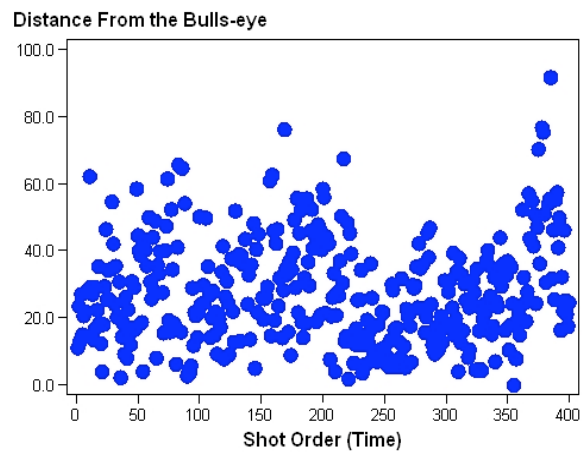

**Figure 4:** Distance from Bulls-eye to shot in order of shooting sequence (time).

In addition, three targets were found to have two bullet holes in them meaning that three shots were fired into the wrong target. These six shots were invalidated because we could not identify which shot corresponded to which hole.

Overall, 20 shots were invalidated for the above mentioned reasons, leaving 380 experimental units to analyze.

## 4. Data Analysis

### 4.1 Response Surface and Exploratory Work

After we finished shooting, we calculated the group centers $(\bar{x}, \bar{y})$ for each experimental level. From these, we obtained the Euclidean distances from the group centers to the individual bullet holes. Now, using our 380 observations, we used the SAS procedure G3D to plot the fitted response surface, shown on the left of Figure 5, and used the SAS GCONTOUR procedure to produce the corresponding contour plot, shown on the right.\

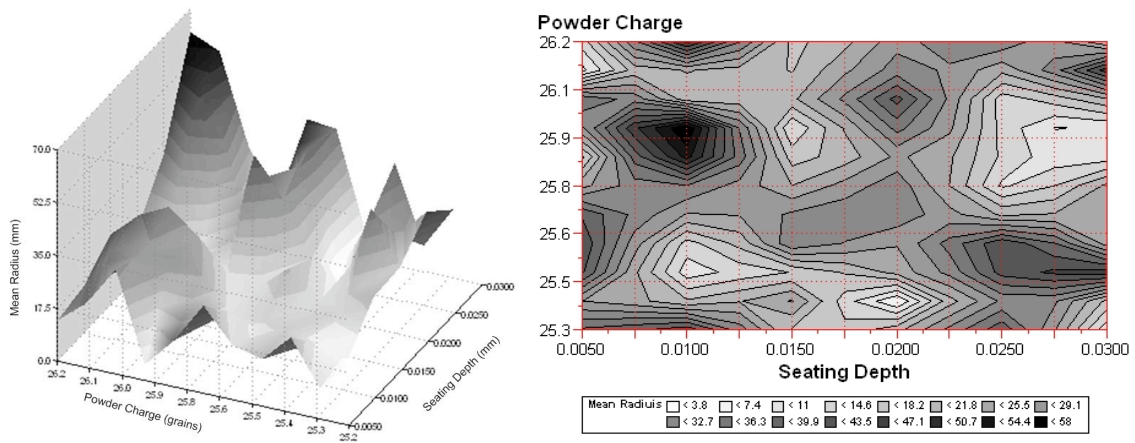

**Figure 5:** Surface Plot and Contour Map of Mean Radius with Respect to Seating Depth and Powder Charge.

As one can see from the response surface and the contour map, there appears to be a fair amount of interaction between our main factors, seating depth (SD) and powder charge (PC). We notice a valley running along the 25.9 grain PC level from 0.015" to the 0.030" SD levels. Another valley runs along the 25.5 grain PC level from 0.005" to the 0.020" SD levels. Both valleys are terminated by peaks, which make a rather symmetrical pattern in the response surface. This response surface gives us an initial idea of what the favourable factor levels might be.

### 4.2 ANOVA Model

We now used the SAS GLM procedure to perform an analysis of variance with seating depth and powder charge as the main factors and lot as our blocking factor. The response variable was the Euclidean distance from the group center for each factor level. The results are summarized in the ANOVA table below, which shows that the overall model was statistically significant, with a *p*-value of 0.0008.

| Table 3. ANOVA Table | | | | | |
|---|---|---|---|---|---|
| Source | DF | Sum of Squares | Mean Square | F Value | Pr > F |
| Model | 62 | 17752.7 | 286.3 | 1.78 | 0.0008 |
| Error | 317 | 51124.8 | 161.3 | | |
| Corrected Total | 379 | 68877.6 | | | |

Table 4 shows the break down per factor. We see that the seating depth by itself and the lot were not statistically significant. Powder charge is significant at the 0.05 level with a p-value of 0.0378. Additionally, the interaction between powder charge and seating depth was very significant with a p-value of 0.0006.

Table 4. ANOVA Results showing p-values for the Class Variables.

| Source | DF | Type I SS | Mean Square | F Value | Pr > F |
|---|---|---|---|---|---|
| lot | 3 | 109.7 | 36.6 | 0.23 | 0.8778 |
| Seating Depth | 5 | 614.9 | 123.0 | 0.76 | 0.5774 |
| Powder Charge | 9 | 2917.1 | 324.1 | 2.01 | 0.0378 |
| Seating Depth * Powder Charge | 45 | 14111.0 | 313.6 | 1.94 | 0.0006 |

This means that powder charge seems to be the dominating factor in determining the optimal cartridge dimensions. Yet, due to the interaction, one should not dismiss the seating depth as an important factor.

### 4.3 ANCOVA Model

Our next step was to perform an analysis of covariance to incorporate the thirteen covariates that we mentioned in section 2.3. We ran the SAS GLM procedure with those covariates included in the model. The results are summarized in Tables 5 and 6 below.

Table 5. ANOVA Table for Analysis of Covariance

| Source | DF | Sum of Squares | Mean Square | F Value | Pr > F |
|---|---|---|---|---|---|
| Model | 75 | 20414.4 | 272.2 | 1.71 | 0.0009 |
| Error | 304 | 48463.2 | 159.4 | | |
| Corrected Total | 379 | 68877.6 | | | |

Table 6. ANCOVA Results showing p-values for the Class Variables and Covariates

| Source | DF | Type I SS | Mean Square | F Value | Pr > F |
|---|---|---|---|---|---|
| lot | 3 | 109.7 | 36.6 | 0.23 | 0.8759 |
| Seating Depth | 5 | 614.9 | 123.0 | 0.77 | 0.5709 |
| Powder Charge | 9 | 2917.1 | 324.1 | 2.03 | 0.0355 |
| Seating Depth * Powder Charge | 45 | 14111.0 | 313.6 | 1.97 | 0.0005 |
| Case Length | 1 | 54.8 | 54.8 | 0.34 | 0.5581 |
| Case Mouth Square | 1 | 189.7 | 189.7 | 1.19 | 0.2763 |
| Case Volume | 1 | 451.5 | 451.5 | 2.83 | 0.0934 |
| Case Weight | 1 | 50.2 | 50.2 | 0.31 | 0.5753 |
| Head Space | 1 | 1.9 | 1.9 | 0.01 | 0.9134 |
| Neck Inner Diameter | 1 | 446.7 | 446.7 | 2.80 | 0.0952 |
| Neck Outer Diameter | 1 | 22.1 | 22.1 | 0.14 | 0.7096 |
| Neck Thickness | 1 | 10.4 | 10.4 | 0.06 | 0.7990 |
| Primer Pocket Depth | 1 | 133.9 | 133.9 | 0.84 | 0.3602 |
| Primer Pocket Diameter | 1 | 19.4 | 19.4 | 0.12 | 0.7278 |
| Primer Weight | 1 | 1167.5 | 1167.5 | 7.32 | 0.0072 |
| Bullet Weight | 1 | 13.4 | 13.4 | 0.08 | 0.7721 |
| Bullet Overall Length | 1 | 100.2 | 100.2 | 0.63 | 0.4285 |

We see from Table 6 that only one of the covariates, the primer weight, is statistically significant. The primer weight is the weight of the primer before being inserted into the primer pocket. The primer weight in our model has two levels, 3.1 grains and 3.2 grains. Since the observed mean radius for a weight of 3.1 grains was smaller than the observed mean radius for 3.2 grains, we conclude that using a 3.1 grain primer produced more accurate results.

### 4.4 Identification of the Minimal Responses

To identify the minimal mean response for the factor-levels, we tabulated and sorted the mean radius, and identified the "best" ones and their corresponding factor-levels. The two lowest mean responses are listed in table 7 along with the largest mean response for comparison.

| Table 7. The Two Minimal Responses and the Maximum Response for Comparison. | | | |
|---|---|---|---|
| Response | Mean Radius (mm) | Seating Depth (inches) | Powder Change (grains) |
| Smallest | 10.62 | 0.030 | 25.9 |
| Second Smallest | 12.60 | 0.005 | 25.8 |
| Largest | 40.97 | 0.030 | 26.2 |

## 5. Summary

To our knowledge, this is the first time that a study of this type has been conducted as a designed and controlled experiment. For the most part, the external conditions were fairly consistent and conducive to this experiment. Any aberrations were carefully accounted for, and the resulting data points were promptly removed. The main results that emerged from this study are: (i) By itself, the seating depth of the bullet does not seem to affect that shooting precision; (ii) The powder charge does seem to affect the shooting precision significantly; (iii) There exists a significant interaction between the seating depth and the powder charge; (iv) Among the various covariates that are associated with the unit-to-unit variation that might have contributed to the variation in the response, only primer weight is significant; and (v) the optimal cartridge dimension for this particular rifle had a seating depth of 0.030 and a powder charge of 25.9 grains. Overall, we believe that our methodology and its implementation to this study are scientifically sound. Therefore, our results and conclusions can serve as valid recommendations for other marksmen.

## Acknowledgements

This research was done as a senior project of the first author, Steven Anderson, as a student at the Metropolitan State College of Denver. Steven Anderson designed the study, performed the shooting and ran the analysis. This senior project was done under the supervision of Dr. Shahar Boneh and Dr. Nels Grevstad, in the college's Department of Mathematical and Computer Sciences. Steven Anderson would like to thank Dr. Shahar Boneh and Dr. Nels Grevstad for advising him in all aspects of this project.